\documentclass[preprint,aps,floats,subeqn,showpacs]{revtex4}

\usepackage{graphicx}
\usepackage{amsmath}
\usepackage{array}
\usepackage{bm}
\usepackage{amssymb}
\usepackage{amsfonts}

\usepackage{epsfig} 
\usepackage[hang,nooneline]{subfigure}

\usepackage{color}

\newcommand{\be}{\begin{equation}}
\newcommand{\ee}{\end{equation}}
\newcommand{\bea}{\begin{eqnarray}}
\newcommand{\eea}{\end{eqnarray}}
\newcommand{\bml}{\begin{mathletters}}
\newcommand{\eml}{\end{mathletters}}

\begin{document}





\title{Q-ball formation at the deconfinement temperature in large-$N_c$ QCD}

\author{Yves Brihaye}
\email[E-mail: ]{yves.brihaye@umons.ac.be}
\affiliation{Service de Physique Th\'{e}orique et Math\'{e}matique,
Universit\'{e} de Mons--UMONS,
Acad\'{e}mie universitaire Wallonie-Bruxelles,
Place du Parc 20, B-7000 Mons, Belgium}

\author{Fabien \surname{Buisseret}}
\email[E-mail: ]{fabien.buisseret@umons.ac.be}
\affiliation{Service de Physique Nucl\'{e}aire et Subnucl\'{e}aire,
Universit\'{e} de Mons--UMONS,
Acad\'{e}mie universitaire Wallonie-Bruxelles,
Place du Parc 20, B-7000 Mons, Belgium; \\ 
Haute Ecole Louvain en Hainaut (HELHa), Chauss\'ee de Binche 159, B-7000 Mons, Belgium}

\date{\today}
\setlength{\footnotesep}{0.5\footnotesep}

\begin{abstract}
The deconfinement phase transition in large-$N_c$ QCD is studied within the framework of an effective Polyakov-loop model, where the potential has a U(1) symmetry originating in the large-$N_c$ limit of a Z$_{N_c}$-symmetric model. At the critical temperature, the shape of the effective potential allows the existence of Q-balls as position-dependent fluctuations of the Polyakov loop. Q-balls with spherical or axial symmetry are numerically obtained from the equations of motion of the effective model under consideration. The physical properties of these non-topological solitons (mass, charge and size) are discussed, as well as their interpretation in terms of spinning ``bubbles", with various shapes, of deconfined matter surrounded by a confined environment.  
\end{abstract}

 \maketitle
 
\section{Introduction}
The phenomenology related to the deconfined phase of hadronic matter -- the quark–gluon plasma -- is nowadays a matter of intense investigations. From a theoretical point of view, the study of gauge theories at finite temperature is a challenging problem, while the QCD matter is or will be studied in heavy-ion collisions at RHIC, SPS, FAIR, and the LHC. Informations and relevant references on the topic can be found for example in \cite{Rev}.

For a given Yang-Mills theory at a nonzero temperature $T$, the Polyakov loop is defined as $L(T, \vec y)= P\, {\rm e}^{i\, g\int^{1/T}_0d\tau A_0(\tau, \vec y)}$, with $A_0$ the temporal component of the Yang-Mills field and $\vec y$ the spatial coordinates. As usual, $P$ is the path-ordering, $g$ is the strong coupling constant and units where $\hbar = c =k_B= 1$ are used. The Polyakov loop is actually such that $\left\langle L(T,\vec y) \right\rangle=0 $ $(\neq 0)$ when the theory is in a (de)confined phase~\cite{Polya}. Since gauge transformations belonging to the center of the gauge algebra only cause $L(T,\vec y)$ to be multiplied by an overall factor, it is tempting to conjecture that the confinement/deconfinement phase transition might be linked to the spontaneous breaking of a global symmetry related to the center of the considered gauge algebra. In the particular case of SU($N_c$), deconfinement might thus be driven by the breaking of a global Z$_{N_c}$ symmetry~\cite{sve82}, with the following dimensionless order parameter
\begin{equation}\label{phi}
\phi=\frac{1}{N_c}{\rm Tr}_c L.
\end{equation}
Note that the color-averaged Polyakov loop $\phi$ will simply be called Polyakov loop in the following. 

It is well known that the thermodynamic properties of pure gauge SU(3) QCD can be studied by resorting to an effective scalar field theory where the potential energy density is Z$_3$-symmetric, with \textit{e.g.} the form~\cite{Pisa} $U\propto -a_2\, |\phi|^2+a_4\, |\phi|^4+a_3(\phi^3+\phi^{*3})$, see also Refs. \cite{Dumitru00,Pisa_app}. It has further been proposed in \cite{sanni05} that, for an arbitrary number of colors, the potential energy density should be of the $Z_{N_c}$-symmetric form $U=a_2\, |\phi |^2+a_4\, |\phi |^4+a_{N_c}\, (\phi^{N_c}+\phi^{*N_c})$. Moreover, the large-$N_c$ scaling and the temperature-dependence of the thermodynamic observables severely constrain the coefficients $a_i$ at large $N_c$, as it has been shown in \cite{Buiss11}. One then gets a potential that becomes U(1)-symmetric in the large-$N_c$ limit. The main results of \cite{Buiss11} concerning the Polyakov-loop effective potential will be summarized in Sec. \ref{PLL}. 

Polyakov-loop-inspired approaches are a precious tool to understand the QCD phase diagram: We refer the interested reader to \textit{e.g.} Refs. \cite{fukupnjl}, where the Polyakov loop is both coupled to quark fields and responsible for an effective potential. A direction that has been less studied so far is to work beyond mean field approximation and let the Polyakov loop fluctuate thanks to a standard kinetic term of the form $\partial_\mu \phi \partial^\mu \phi^*$: Dynamical effects like hadronization through decay of the Polyakov loop \cite{Sca01,Sca02} or formation of plasma bubbles in heavy ion collisions \cite{Gupta10} can then be studied. The plasma bubbles found in the above references are actually static and spherically symmetric solutions of the equation of motion associated to a given Polyakov-loop Lagrangian; they can be interpreted as a sphere of deconfined matter surrounded by a confined medium. One of the main goals of the present paper is to go a step further in this direction and show that \textit{spinning} plasma bubbles, and more generally bubbles with axial symmetry, may also exist. Such bubbles can reasonably be imagined to form in non central heavy-ion collisions, where the incident nulcei have a relative angular momentum. 

It should be pointed out that there exist alternative descriptions of the pure gauge theory which make use of the transverse particle concept, see e.g. \cite{Meisinger:2003id,Sasaki:2012bi,Ruggieri:2012ny}. Descriptions of the physical phenomenon based  on the AdS/QCD correspondance principle are also available, see namely \cite{Kajantie:2006hv,Bigazzi:2011db,Kelley:2011ds}.

The paper is organized as follows~:
The model we use is presented in Sec. \ref{Model}. It is directly inspired from the large-$N_c$, U(1) symmetric, Polyakov loop potential presented in \cite{Buiss11}. The advantage of a U(1) symmetry is that it allows in principle the existence of Q-balls. In general, Q-balls refer to regular, finite mass and localized classical solutions of a 
complex, self-interacting scalar field theory \cite{coleman,lee_pang}; they form  non-topological solitons of the underlying equations.
In the present context, these solitons will come out as natural solutions of the Polyakov loop equation of motion. Q-balls with spherical and axial symmetry will be found by numerical resolution of the Polyakov-loop equations of motion in Secs. \ref{Sphe} and \ref{Axial} respectively. Their interpretation in terms of bubbles of deconfined plasma will be discussed, as well as their main physical properties. The various solutions obtained are compared in Sec. \ref{compar}, while a discussion of the possible extensions of the present results to finite-$N_c$ cases and concluding comments are given in Secs. \ref{finite} and \ref{Conclu}.

\section{Polyakov-loop Lagrangian}\label{PLL}

According the arguments exposed in the introduction, an effective Lagrangian for a finite-temperature SU($N_c$) Yang-Mills theory should involve the Polyakov loop (\ref{phi}), which is seen as a complex scalar field, as well as a  Z$_{N_c}$-symmetric potential, denoted $V_g$. In order to mimic at best known results in Yang-Mills theory, $V_g$ should be such that:
\begin{itemize}
\item The pressure $p_g=-{\rm min}_{\phi}(V_g)$ is proportional to $N_c^2\, T^4$ at large $N_c$ and $T$ in order to recover asymptotically the Stefan-Boltzmann limit for a free gluon gas.
\item The value of the Polyakov loop minimizing $V_g$ is $N_c$-independent at the dominant order \cite{make}. Numerically, $|\phi_0|=0$ in the confined phase, $>0$ in the deconfined phase, and tends toward unity at very large $T$. 
\item There exists a critical temperature $T_c$ above which the absolute minimum is nonzero, so that one has a first-order phase transition. $T_c$ has to be seen as a typical value for the deconfinement temperature in SU($N_c$) Yang-Mills theory since the deconfinement temperature appears to be $N_c$-independent up to corrections in $1/N_c^2$~\cite{TcN}. According to previously obtained results, the norm of the Polyakov loop when $T\rightarrow T_c^+$ may be found between 0.4 and 0.6 \cite{gupta07,braun}.
\end{itemize} 

\begin{figure}[t]
\includegraphics[width=10cm]{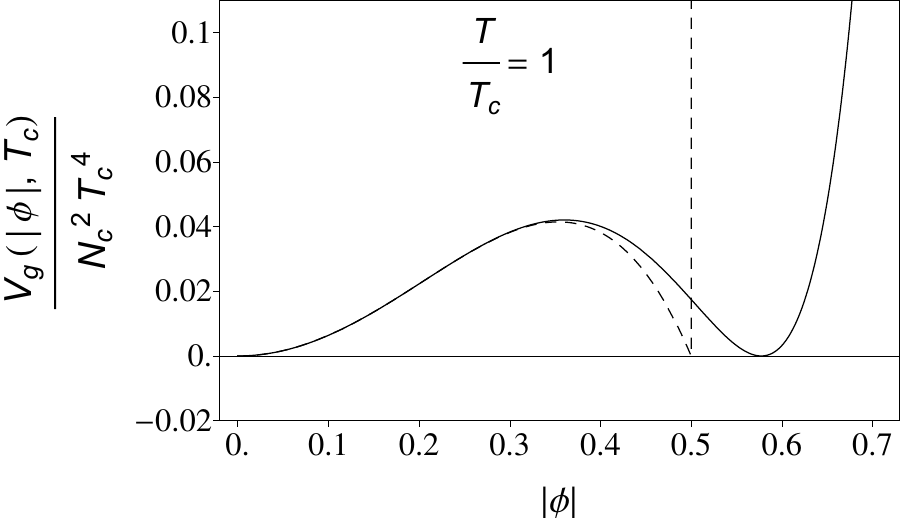}
\caption{\label{fig1}
Polyakov-loop potential $V_g(|\phi |_c)$ versus $|\phi |$ in the large-$N_c$ limit (dashed line); $V_g$ is given by (\ref{ldef}) and (\ref{V3}). The potential used in our numerical calculations, that is (\ref{V4}), is also plotted for comparison with $\beta=10$ and $C=C_{10}=27$ according to (\ref{Cdef}) (solid line).}
\end{figure}

It has been shown in \cite{Buiss11} that the potential
\begin{equation}\label{V2}
V_g=N_c^2 T^4\, a(T)\left[ |\phi|^2-4|\phi|^4+\frac{l(T)^{2-N_c}}{N_c}[8 l(T)^2-1](\phi^{N_c}+\phi^{*N_c})\right],
\end{equation}
with 
\begin{equation}\label{ldef}
l(T)=0.74-0.26\, \tanh\left[2.10\left(\frac{T_c}{T}\right)^3-0.60\frac{T}{T_c}\right] \ {\rm and} \  a(T)=\frac{1}{l(T)^4}\left(\frac{\pi^2}{135} -\frac{0.029}{\ln(T/T_c+1.5)}\right),
\end{equation}
has all the required features and is moreover such that it leads to numerical values of the pressure and of the Polyakov loop in excellent agreement with recent lattice data \cite{gupta07,pane09}, both for $N_c=3$ and in the large-$N_c$ limit. This last limit is particularly interesting for our purpose since, when $N_c\rightarrow\infty$, $V_g$ becomes U(1)-symmetric and reads \cite{Buiss11}
\begin{eqnarray}\label{V3}
V_g(|\phi |, T)=& N_c^2 T^4 \, a(T)\, |\phi |^2(1-4 |\phi |^2)  &\qquad |\phi |\leq l(T),\nonumber\\
& +\infty &\qquad |\phi |>l(T).
\end{eqnarray} 
It is worth mentioning that $l(T_c)=1/2$, $a(T_c)=2/3$, and that, as expected in the case of a first-order phase transition, the two minima are degenerate at the critical temperature: $V_g(0,T_c)=V_g(1/2,T_c)=0$. The shape of the potential at $T=T_c$ is shown in Fig. \ref{fig1}; it is the typical kind of shape for which soliton-like solutions exist, as we will show in the following. 

Potential (\ref{V3}), becoming infinite once $ |\phi |$ is large than some $T$-dependent value, is not very convenient for numerical applications such as those that are to be performed. Instead of using this last form, we will rather introduce the following potential
\begin{eqnarray}\label{V4}
V_g(|\phi |, T_c)=&\frac{2}{3}  N_c^2 T_c^4 \, (|\phi |^2-4 |\phi |^4+C |\phi |^\beta)  ,
\end{eqnarray} 
which can be seen as a more tractable approximation of (\ref{V3}) at $T=T_c$, with a term in $|\phi |^\beta$ introduced to mimic the ``wall" of (\ref{V3}). Generic values of $C$ can be considered, but the main emphasis will be put on the particular choice 
 \be\label{Cdef}
C=C_\beta \equiv \frac{4 |\phi_\beta|^4 - |\phi_\beta|^2}{|\phi_\beta|^\beta}, \ {\rm where}\    |\phi_\beta| = \sqrt{\frac{\beta-2}{4\beta-16}}.
 \ee
 It is such that potential (\ref{V4}) is still zero in its two degenerate minima, namely $|\phi|=0$ and $|\phi|=|\phi_\beta|$. This second minimum tends toward $1/2$ when $\beta$ is arbitrarily large: (\ref{V3}) is then recovered as the limit of (\ref{V4}) when $\beta$ tends toward infinity. The potential (\ref{V4}) is shown in Fig. \ref{fig1} for $\beta=10$ and $C=C_{10}=27$ according to (\ref{Cdef}): One sees that, though not being an extremely accurate approximation of (\ref{V3}) when $ |\phi |>0.5$, it shares the same structure, and will presumably lead to qualitatively similar results, which is satisfactory for our mostly exploratory purpose. Note that $|\phi_{10}|=0.58$ instead of $1/2$, but this is still acceptable if one considers values previously found in computations of the critical value of the Polyakov loop \cite{braun}. 

It sould be pointed out that the replacement of the potential (\ref{V2}) by (\ref{V4}) 
introduces an extra conserved quantity related to the continuous symmetry (for instance the charge $Q$ defined in the following). The
consequence of this will be commented in due course. 
 
\section{The model and the equations}\label{Model}

\subsection{Reduced Lagrangian}

We now consider that the Polyakov loop $\phi$ is a dynamical complex scalar field. According to the suggestion of \textit{e.g.} Ref. \cite{Dumitru00}, we take a kinetic part of the form $N^2_c T^2 \partial_\mu \phi \partial^\mu \phi^* / \lambda$, which has both the correct energy dimensions and the expected $N_c$-scaling. $\lambda$ is the 't Hooft coupling. The starting Lagrangian, valid at $T=T_c$, is thus given by
\be
           {\cal L}_{phys} = \frac{N^2_c T_c^2}{\lambda} \partial_\mu \phi \partial^\mu \phi^* -\frac{2}{3}  N_c^2 T_c^4 \, (|\phi |^2-4 |\phi |^4+C |\phi |^\beta)  ,
\ee
where $\phi=\phi(y^\mu)$, $y^\mu$ being the spacetime coordinates. It is natural to further define dimensionless variables
$x^{\mu}$ related to the original (physical) ones by
\be
          y^{\mu} =  l_{phys} \ x^{\mu},\ {\rm with}\ l_{phys}=\frac{1}{T_c} \sqrt{\frac{3}{2 \lambda}} ,
\ee
so that the above Lagrangian can be replaced by
\be \label{lag0}
          {\cal L}=\frac{ {\cal L}_{phys}}{(2/3)N_c^2 T_c^4} = \partial_\mu \phi \partial^\mu \phi^* -V(|\phi |) ,
\ee
where $\phi=\phi(x^\mu)$ and 
\be \label{V0}
V(|\phi |)=|\phi |^2-4 |\phi |^4+C |\phi |^\beta .
\ee
The corresponding classical equations of motion read
\be
      \partial_\mu \partial^\mu \phi = \frac{\partial V}{\partial \phi^*} 
      =  \phi\, ( 1 - 8 |\phi|^2 + \frac{\beta C}{2} |\phi |^{\beta-2})  , 
\ee
plus the complex conjugated equation.
\subsection{The ansatz for Q-balls and conserved quantities}
 The standard way to construct Q-balls consists in looking for solutions of the form 
 \be
 \label{ansatz0}
 \phi(x^\mu) = \exp(i\omega x^0) \psi(\vec x),
 \ee
where $\psi(\vec x)$ is a function of the space variables. The real parameter $\omega$ constitutes an essential characterisation of the stationary
solution. As will be soon discussed, solitons only exist for $\omega$ taking values in a finite interval.

Two types of solutions will be searched for: Spherically- or axially-symmetric ones. Spherically symmetric solutions correspond to further assuming $\psi(\vec x)=\chi(r)$, where $r$ is the radius in spherical coordinates. Axially symmetric solutions correspond to the ansatz $\psi(\vec x)=\exp(i k \varphi) \chi(r, \theta) $, where $\{r,\varphi,\theta\}$ denote the standard spherical coordinates and where $k$ is an integer. We remark that $\chi$ is assumed to be a real function.

The solutions we will build can be characterised by their energy $M$  and by a dimensionless conserved charge $Q$, respectively defined by 
\be \label{Mdef}
           M =M_{phys} \int d^3 x \ T_{00}
\ee
and
\be \label{Qdef}
            Q = 2 \omega \int d^3 x \ |\phi |^2 .
\ee
The temporal component of the energy-momentum tensor represents the energy density, given by
\be
T_{00}=\omega^2 |\phi |^2 + \vec\nabla \phi\cdot \vec\nabla \phi^* +V(|\phi |).
\ee
The natural mass scale introduced above reads  
\be \label{Mphys}
              M_{phys} = \sqrt{\frac{3}{2}} \frac{N_c^2 T_c}{\lambda^{3/2}}.
\ee

The conserved charge $Q$ finds its origin in the U(1)-symmetry of the considered lagrangian, leading to a conserved Noether current
of the form  $J_{\mu} = i( \phi \partial_{\mu} \phi^*  - \phi^* \partial_{\mu} \phi)$, $Q$ being the the space integral of $J_0$. 
Note that the axially symmetric solutions having  $k\neq 0$ are spinning Q-balls whose angular momentum $J$ is related to the charge $Q$ according to $J=k Q$ \cite{volkov}. This relation is specific to Q-balls: the solutions constructed with  the potential (\ref{V2})
could still have a conserved angular momentum while $Q$ would be meaningless.

Let us point out that all the masses and lengths to be plotted below are expressed in units of $M_{phys}$ and $l_{phys}$ respectively.

\subsection{Physical quantities}

At this point we have to stress that the proposed model does not intend to provide an accurate description of ``realistic" plasma bubbles that might form in heavy ion collisions. It is rather a first step to explore the different kinds of soliton-like solutions of a typical Polyakov-loop-inspired model of the QCD phase transition. Nevertheless, it has been shown in Ref. \cite{Buiss11} that our model correctly reproduces the QCD equation of state in the large-$N_c$ limit computed on the lattice \cite{pane09}, and leads to a commonly accepted shape for the QCD phase diagram once coupled to quarks. Therefore we think that the Q-balls we will obtain have something to tell, at least qualitatively, on non-trivial phenomena that might arise at the deconfinement temperature. 

It is worth estimating here the physical mass and length used in the model thanks to some known numbers concerning the quark-gluon plasma at $N_c=3$. First, a typical value for the deconfinement temperature in QCD is $T_c=0.2$ GeV \cite{Rev}. Second, the 't Hooft coupling can be expressed as a function of the strong coupling constant $\alpha_s$ through $\lambda=4N_c\pi\alpha_s$. A way to estimate $\alpha_s$ is to note that the short-range part of the static interaction between a quark and an antiquark scales as $-(4/3)\alpha_s/r$ , at least from $T=0$ to $T_c$. Recent lattice studies, performed at $N_c=3$ and with 2 light quark flavors, favor $\alpha_s=0.2$ up to $T=T_c$ \cite{kacz}, that is the value we retain here. We are then in position to estimate that, at $N_c=3$, 
\be 
l_{phys}=2.2\ {\rm GeV}^{-1}=0.44\ {\rm fm},\quad M_{phys}=0.107\ {\rm GeV}.
\ee

\section{Spherically symmetric solutions}\label{Sphe}
Starting from Lagrangian (\ref{V0}), the field equation reduces, in the spherically symmetric case, to
the differential equation~:
\be\label{eom1}
      \frac{d^2 \chi}{dr^2} + \frac{2}{r} \frac{d \chi}{d r} = \chi (1 - \omega^2  - 8 \chi^2 + \frac{\beta C}{2}  \chi^{\beta-2})
\ee
which has to be solved for $r \in [0,\infty]$. The regularity of the solution at the origin implies $d\chi(0)/dr=0$, the finiteness of the energy and the charge
impose $\chi(\infty)=0$. These conditions specify the conditions at the boundary. 

 Adapting the results of \cite{volkov} to our potential, it can be shown that Q-balls with spherical symmetry exist for
 \be
            \omega_\beta^2 \leq \omega^2 \leq 1  \ \ {\rm with} \ \ 
            \omega_\beta^2 = {\rm min} \{0 , 1 - \frac{4\beta-16}{\beta-2} \left(\frac{8}{C\, (\beta-2)}\right)^{\frac{2}{\beta-2}}  \} 
 \ee
In particular, $\omega_\beta = 0$ if $C \leq C_\beta$. 
Only for $C > C_\beta$ the lower bound is non zero.
The case mostly studied in \cite{volkov, kkl} corresponds (up to an appropriate
renormalisation of the field and of the radial variable) to $\beta=6$ and $C=44/10$ 
so solutions exist for $1/11 < \omega^2 < 1$. 
In the present work, we are mostly interested in the case $C=C_\beta$ for which Q-balls exist for $\omega\in[0,1]$. 
For $\beta=6$, \textit{i.e.} the most commonly used value in studies devoted to Q-balls, we have $C_6 = 4$.

\begin{figure}[t]
    \subfigure
    {\label{soliton_a0_e2}\includegraphics[scale=0.55]{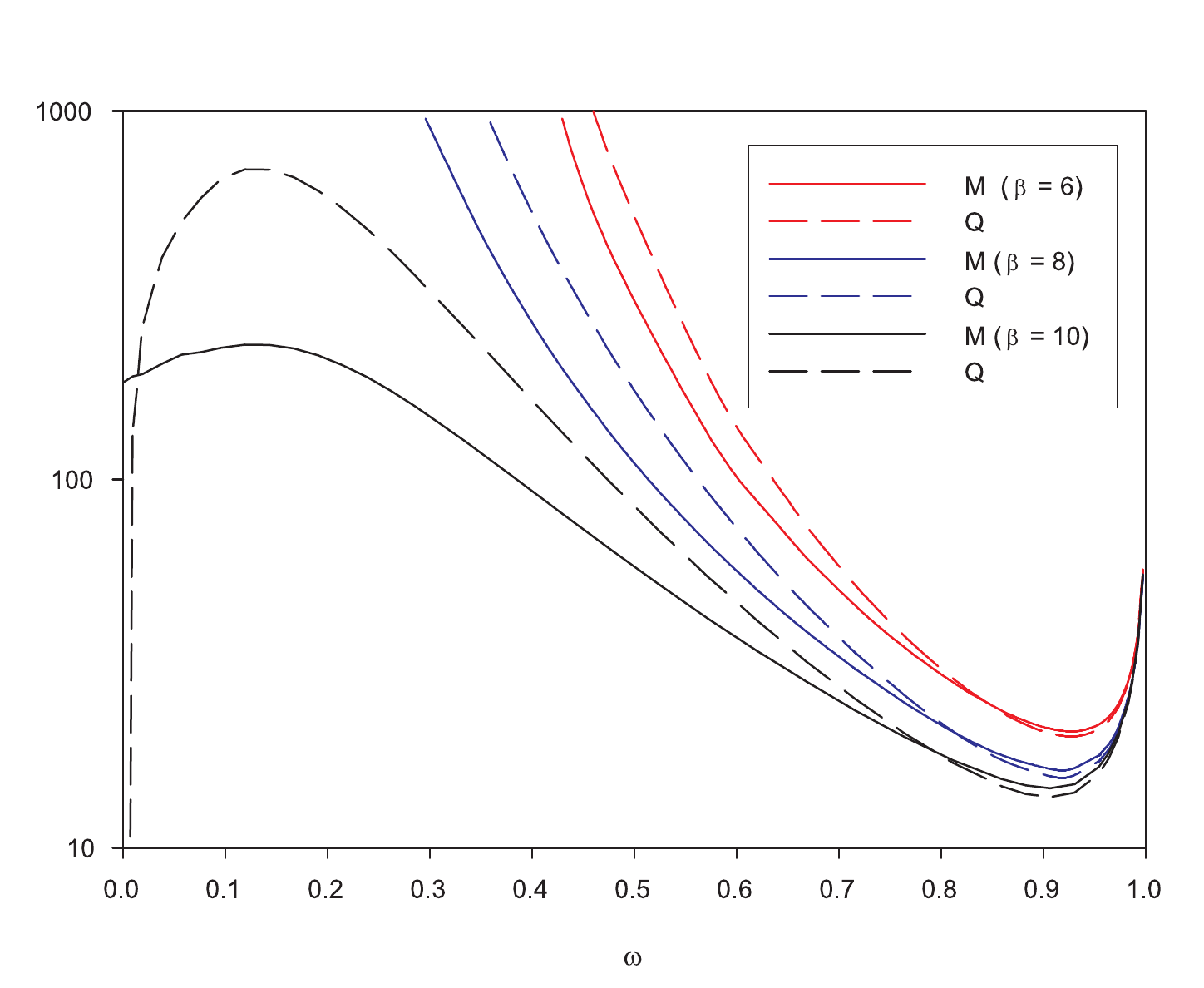}}
    \subfigure
    {\label{soliton_mpsi_e2}\includegraphics[scale=0.55]{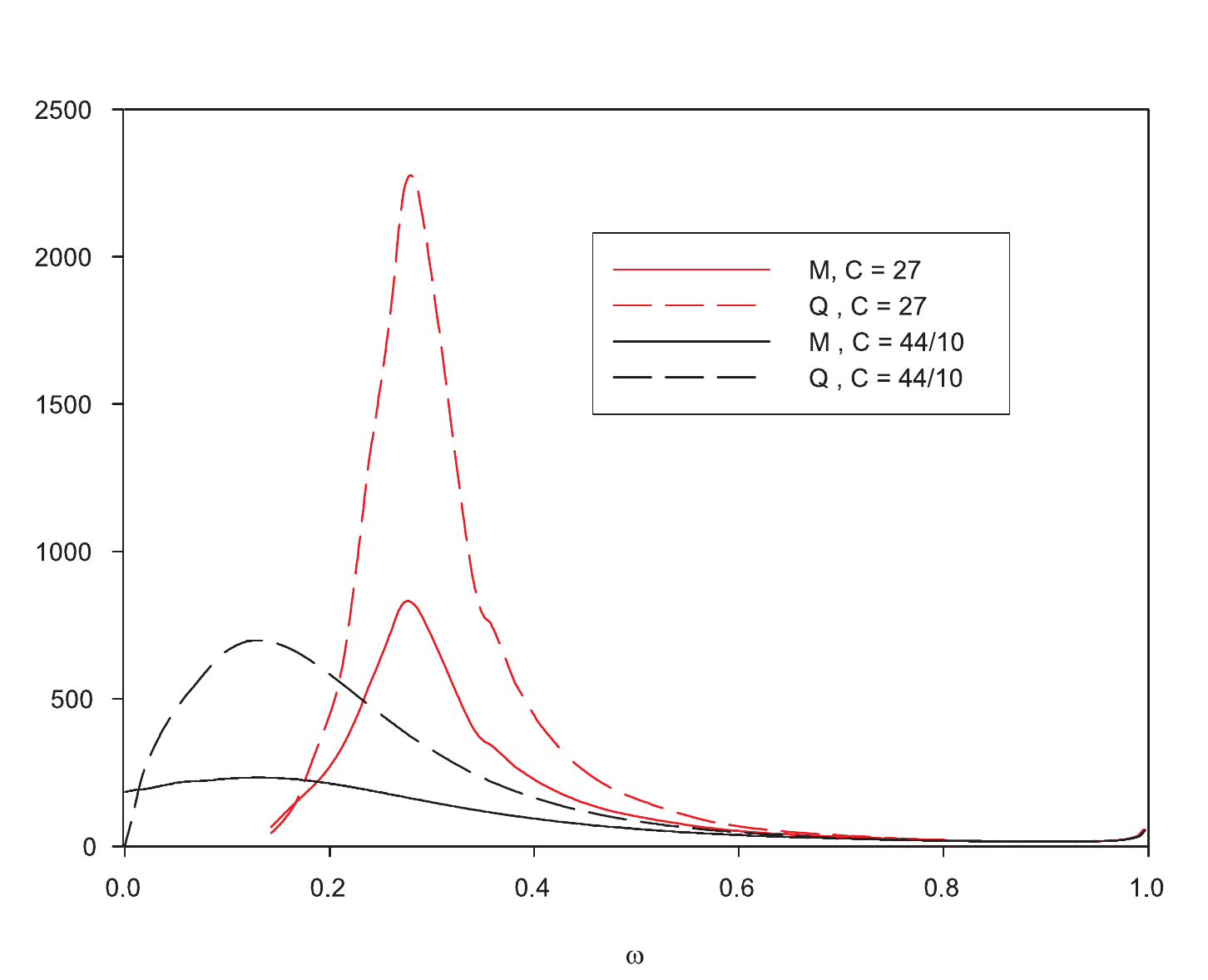}}
   \caption{Left: Mass (solid lines) and charge (dashed lines) of spherically symmetric Q-balls for $C=44/10$ and (from top to bottom) $\beta=6$ ,$8$, $10$. 
            Right: Idem for $\beta=10$ and (from top to bottom) $C=27$, $44/10$.
         }
\label{comparaison} 
 \end{figure}

In order to calibrate our solutions with cases studied in the literature, we have first
solved the equations for $C=44/10$ and several values of $\beta$. The mass $M$ and charge $Q$ of these
solutions are reported as functions of $\omega$ on Fig. \ref{comparaison} (left side).
On the right side, the same quantities are reported for $N=10$ and two values of $C$~:
the value $C=44/10$ corresponds to Ref.\cite{volkov}, and $C=27=C_{10}$ is the case under consideration for this paper. We have checked that we recover indeed previously obtained solutions for the aforementioned values of the parameters. 

It is worth saying that Eq. (\ref{eom1}) has, to our knowledge, no analytic solution for arbitrary values of $\beta$ and has to be
solved numerically. We exclude of our considerations the trivial case $C=\beta=4$, actually leading to a free Schr\"odinger equation for $\chi$. It should be pointed out that even a numerical approach is not obvious because
the boundary conditions are compatible with the trivial solution $\chi(r)=0$. As a consequence,
obtaining a non trivial numerical solution requires a starting guess which is reasonably close to the desired
solution.  We used
 a collocation method for boundary-value ordinary
differential equations, equipped with an adaptive mesh selection procedure \cite{colsys}.
Our solution were constructed with a relative error of order $10^{-8}$.

\begin{figure}[t]

\includegraphics[width=10cm]{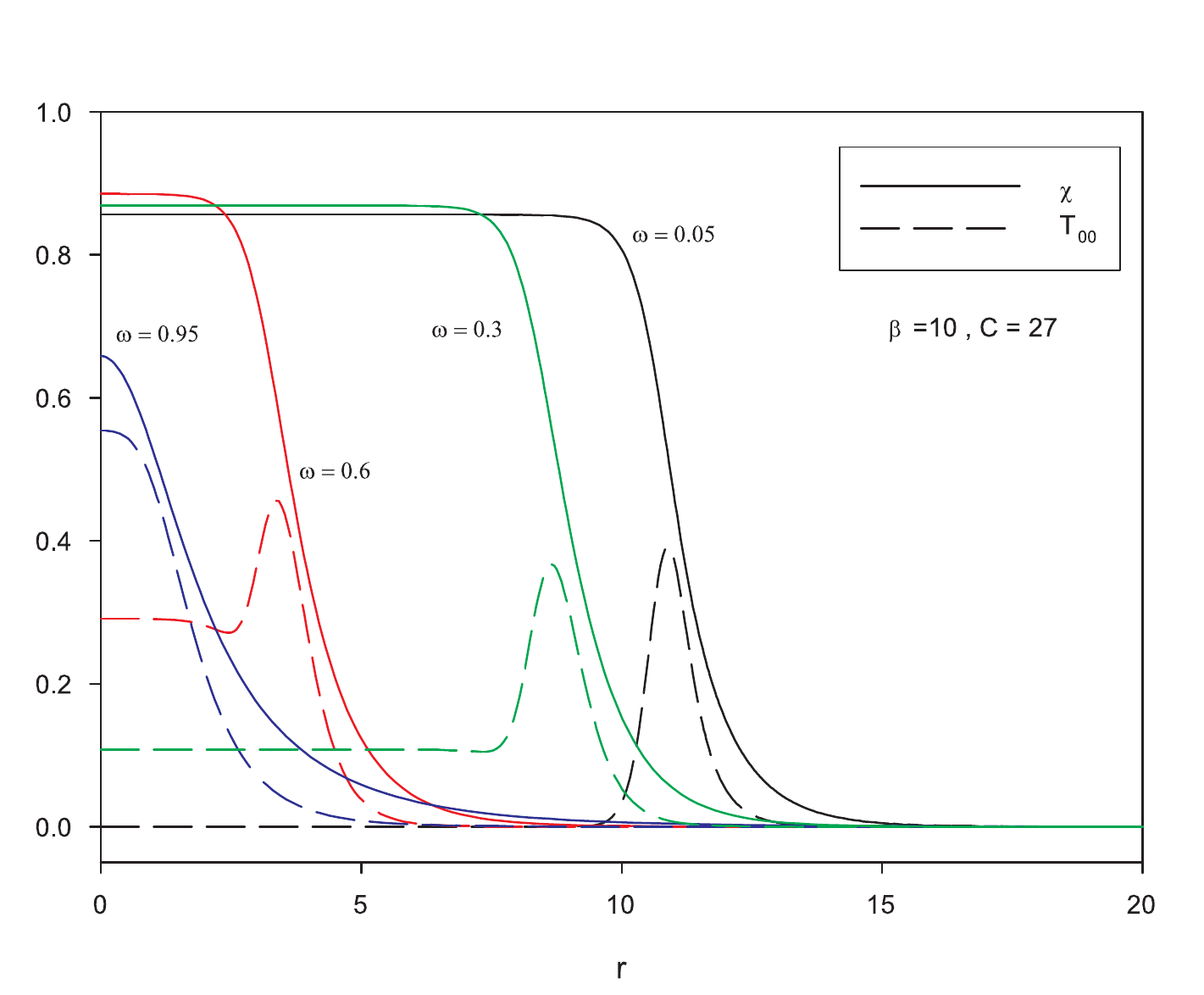}
\caption{\label{profile}
Profile of $\chi(r)$ (solid lines) and of the energy density (dashed lines) in the spherically symmetric case  for $\beta=10$, $C=C_{10}=27$ and (from left to right) $\omega=0.95$, $0.6$, $0.3$, $0.05$.}
\end{figure}

In the following, we focus on the case $\beta =10$ and $C=C_{10}=27$. From a technical point of view, the numerical construction of the solutions, especially the axially symmetric ones, becomes more involved while increasing $\beta$; that is why no values of $\beta$ large than 10 will be used. Nevertheless
we think that the case $\beta = 10$ already captures most of the qualitative features of the limit $\beta \rightarrow \infty$, that corresponds to our original Lagrangian. 
Solutions corresponding to $\omega = 0.1$, $0.3$, $0.6$, $0.95$ are plotted in Fig. \ref{profile} in order to show the behaviour of $\chi(r)$ and $T_{00}(r)$ for various values of $\omega$ lying in the range $[0,1]$.  

For small values of $\omega$, the minimum of the effective potential is deep
and the scalar field ``likes" to stay close to this minimum: It can be observed in Fig. \ref{profile} that, confirming this observation, our numerical results demonstrate that
 the radial function $\chi(r)$ remains practically constant 
(\textit{i.e.} $\chi(r) \sim \chi(0) \equiv\vert \phi_\beta\vert$) 
inside a large sphere centred at the origin. Then it brutally decreases to the the asymptotic value $\chi(r \to \infty) = 0$, forming a thin wall.  Accordingly the energy density presents a plateau for $T_{00} = \omega^2 \phi_\beta^2$  inside the sphere,
 then reaches a maximum on a spherical shell corresponding to the wall and finally decreases to zero. These 
 features appear on Fig. \ref{profile}. The scenario is quite different for $\omega$ close to unity. Here the effective potential's
minimum is not deep anymore and the scalar field quickly reaches its asymptotic value as shown in Fig. \ref{profile}: This is the thick-wall limit.

A question that is worth asking is: Does the Q-balls we have found describe some deconfined sphere -- or spherical shell -- surrounded by a confined environment? Although the Polyakov loop is a straightforward indicator of confinement in a mean-field treatment, this is less obvious in the present framework, where it is never exactly zero, unless asymptotically by construction. We find that a good criterion could be instead given by the value of the energy density. According to the recent lattice computations of the pure gauge QCD equation of state given in \cite{pane09}, one can estimate that the deconfined phase is characterised by energy densities such that $\epsilon\geq 0.35 N^2_c\, T^4_c$. By recalling that $\epsilon=T_{00} M_{phys}/l_{phys}^3$ , we can translate the criterion $\epsilon\geq 0.35 $ into
 \be \label{conf}
 T_{00}\geq 0.45.
\ee
This last inequality can be used in the following as a criterion to seprate the confined regions from the deconfined ones in the solutions we find. A glance at Fig. \ref{profile} shows that only Q-balls with $\omega>0.6$ show deconfined regions according to the above criterion. Only these Q-balls could then be interpreted as ``plasma balls" or ``plasma shells", while the others would just be fluctuations of the Polyakov loop in a completely confined region of space. As it can be observed in the left panel of Fig. \ref{comparaison}, Q-balls with $\omega$ around 0.9 not only contain a deconfined part, but are also the lightest ones that can be obtained. The Q-ball with minimal mass is reached for $\omega=0.88$, with a charge $Q=13.7$ and a mass $M=14.7$ (1.57 GeV). In the thick-wall limit, that is for $\omega=1$, one has finally $Q=55.0$ and $M=55.2$ (5.87 GeV). We mention finally that the Q-ball with $\omega=0.6$ has a charge $Q=44.5$ and a mass $M=35.5$ (3.78 GeV).

\section{Axially symmetric solutions}\label{Axial}

Let us now turn to axially symmetric solutions. To our knowledge, they have never been obtained so far within the framework of a Polyakov-loop model or, more technically, in a Q-ball study using the value $\beta=10$. The field equation reads in this case
\be
\label{pde}
      \frac{\partial^2 \chi}{\partial r^2}  
      + \frac{2}{r} \frac{\partial \chi}{\partial r} 
      + \frac{1}{r^2} \frac{\partial^2 \chi}{\partial \theta^2} 
      + \frac{\cos \theta }{r^2 \sin \theta} \frac{\partial \chi}{\partial \theta} 
      - \frac{k^2 }{r^2 \sin^2 \theta}\chi
      = \chi (1 - \omega^2  - 8 \chi^2 + \frac{\beta C}{2} \chi^{\beta-2}),
\ee
and has to be solved in principle for $r \in [0,\infty]$ and $\theta \in [0, \pi]$. The above equation is notoriously difficult to solve because it is an elliptic non-linear
 partial differential equation. 
 It has to be completed by appropriate boundary conditions
 which are given below. However, rendering the numerical construction more difficult,
  the different sets of conditions are  compatible with the trivial solution $\chi(r,\theta)=0$. 
Several solutions of equations of the type above were constructed in different contexts, see e.g. 
\cite{Arodz:2009ye,Brihaye:2009dx,Campanelli:2009su}.
Comparisons of these solutions with ours constitutes a useful crosscheck of our numerical method. 
For our problem,  the integration was performed numerically by a routine \cite{fidisol} based on the Newton-Raphson method. Concerning the angular dependence of the solutions, we will focus on Q-balls which are even or odd
under the reflection $\theta \to \pi-\theta$. Accordingly the interval of
the variable $\theta$ can  be limited to $\theta \in [0, \pi/2]$. 
In addition to their mass and charge, the solutions with $k>0$ are further characterized by an
 angular momentum given by $J=k Q$ (see \cite{volkov} for details). 

\subsection{Spinning solutions}
\begin{figure}[ht]
  \subfigure{\includegraphics[width=7cm]{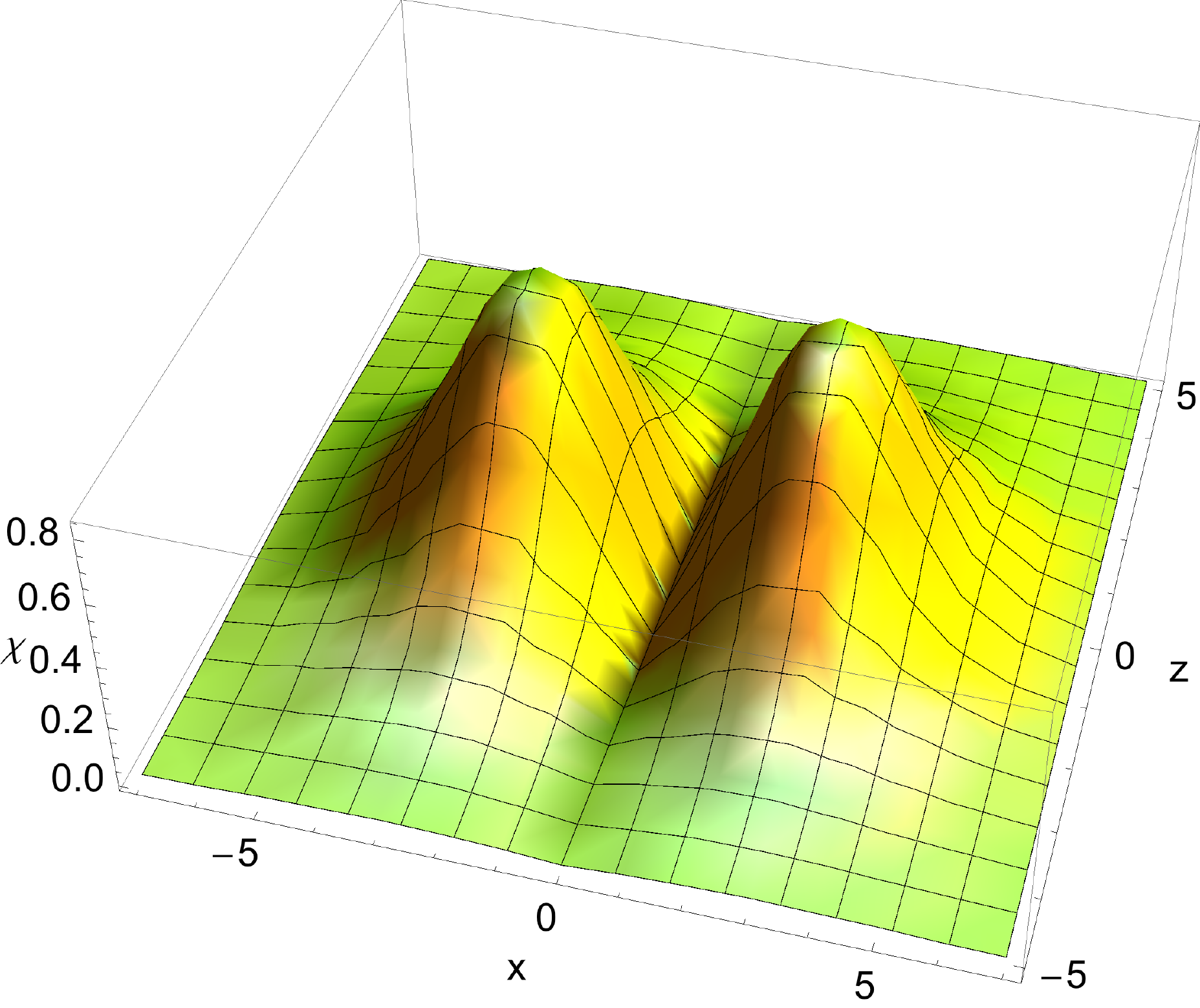}}
    \subfigure{\includegraphics[width=7cm]{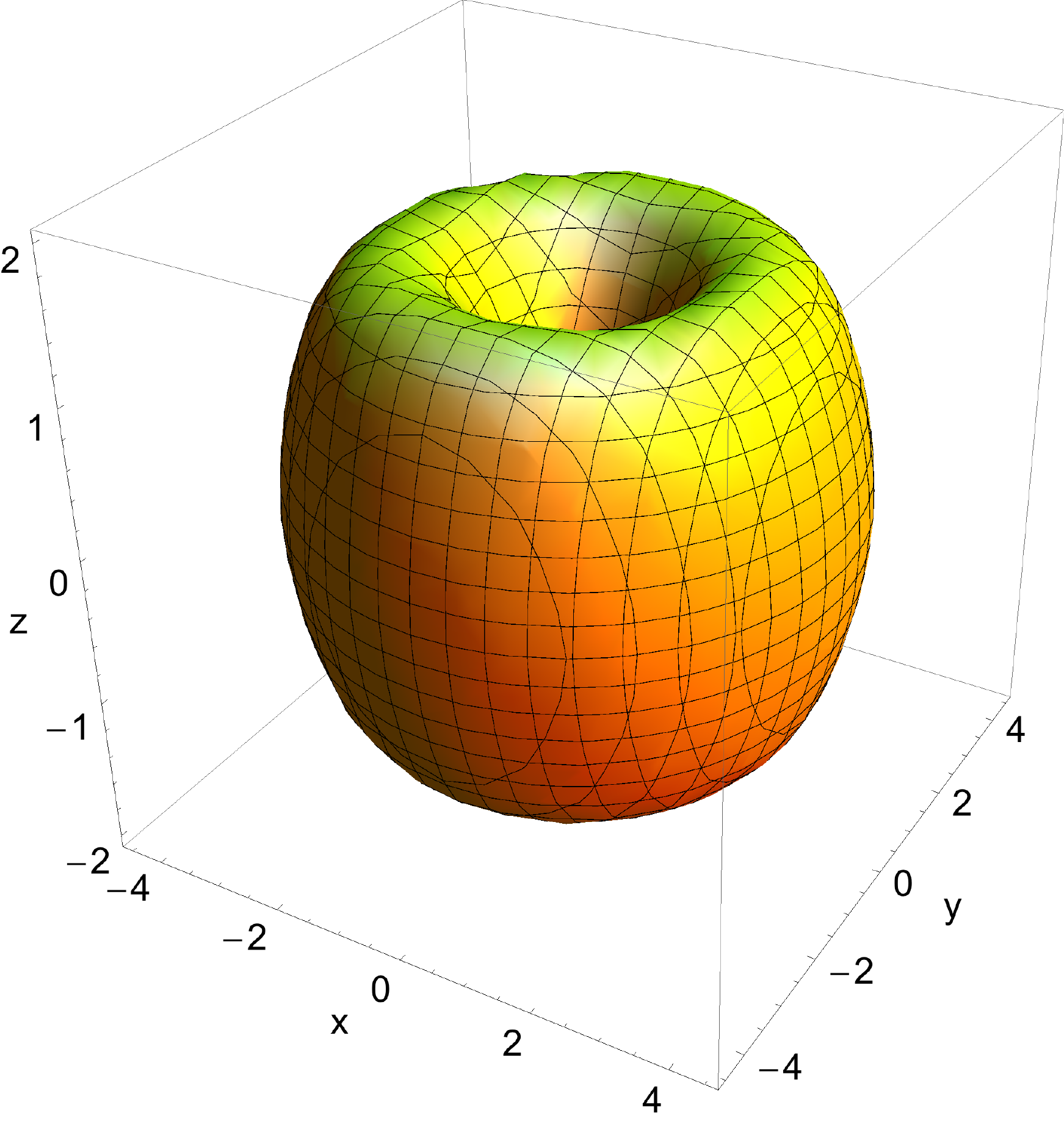}}
\caption{\label{contour_11}
Left: Plot of the scalar field function in the $\phi=0$ plane for the spinning Q-balls with boundary conditions (\ref{bound1}) corresponding to $\beta=10$, $C=C_{10}=27$ and  $\omega=0.8$. Right: Plot of the region where $T_{00}\geq 0.45$ for the same solution. Note that $\{ x,y,z \}$ denote standard cartesian coordinates. }
\end{figure}

Setting $k=1$, Eq. (\ref{pde}) can be solved by imposing the boundary conditions
\be\label{bound1}
      \chi(0, \theta) = 0 \ \ , \ \  \chi(\infty, \theta) = 0 \ \ , \ \ \chi(r, \theta=0) = 0 \ \ , \ \
      \partial_{\theta} \chi(r, \theta=\pi/2) = 0. 
\ee
Solutions of this type are even under the reflexion $\theta \to \pi - \theta$ and have angular momentum $J=Q$. Such Q-balls have been first obtained in \cite{volkov} for $\beta$=6; they can be deformed to the case of our potential. As an illustration, plots of the function $\chi(r,\theta)$ and of the region where the energy density satisfies to the criterion (\ref{conf}) are presented in Fig. \ref{contour_11} for the solution  $\omega = 0.8$ . It is readily seen that the spinning Q-ball found by applying the boundary conditions (\ref{bound1}) has a toric shape, winding and spinning around the $z$-axis. Such a Q-ball may be seen as a ``plasma ring": A spinning torus of deconfined matter. Note that all the values of $\omega$ for which we have found a solution show a similar toric shape and a deconfined part.

\begin{figure}[ht]
    \subfigure
    {\label{soliton_a0_e2}\includegraphics[scale=0.5]{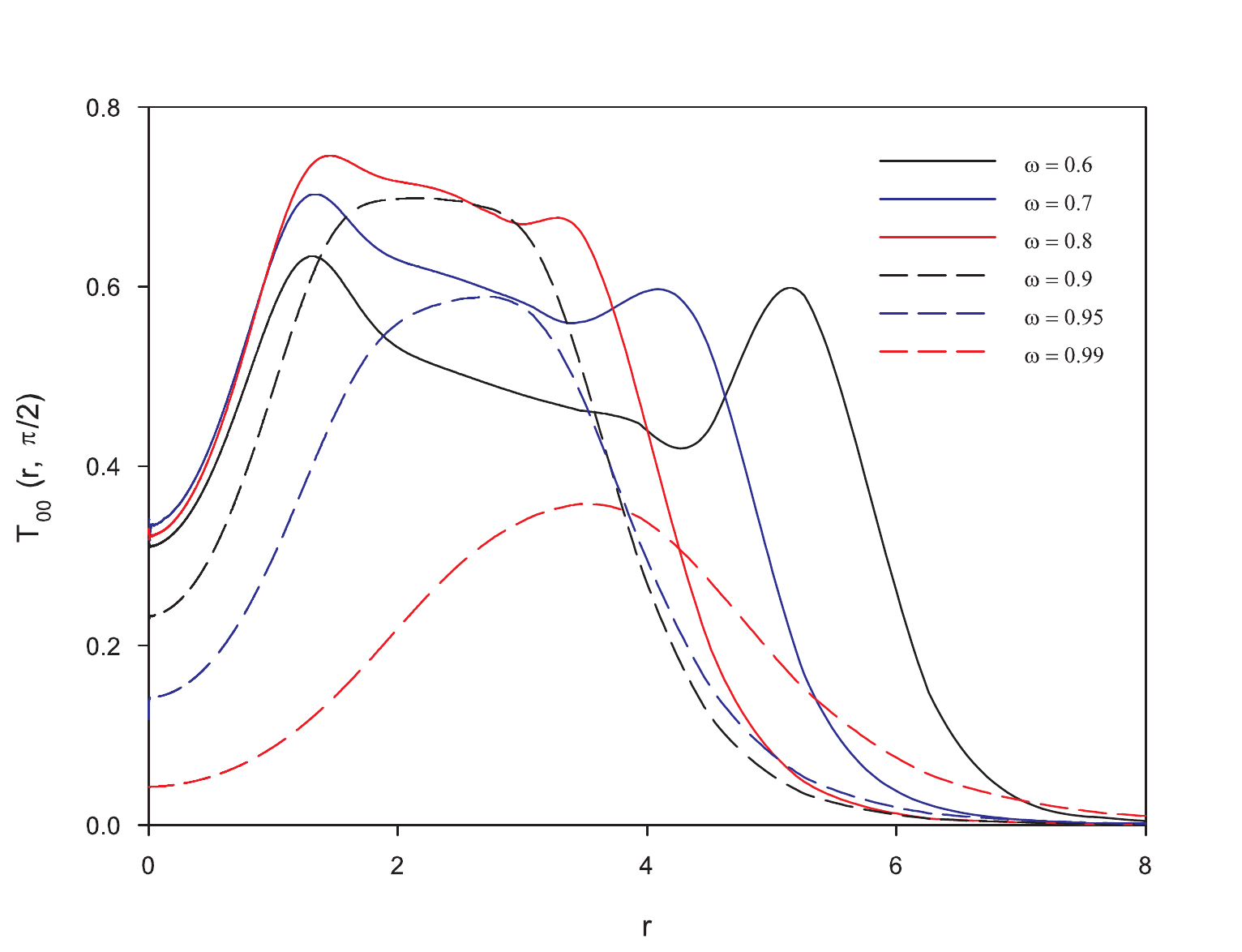}}
    \subfigure
    {\label{soliton_mpsi_e2}\includegraphics[scale=0.5]{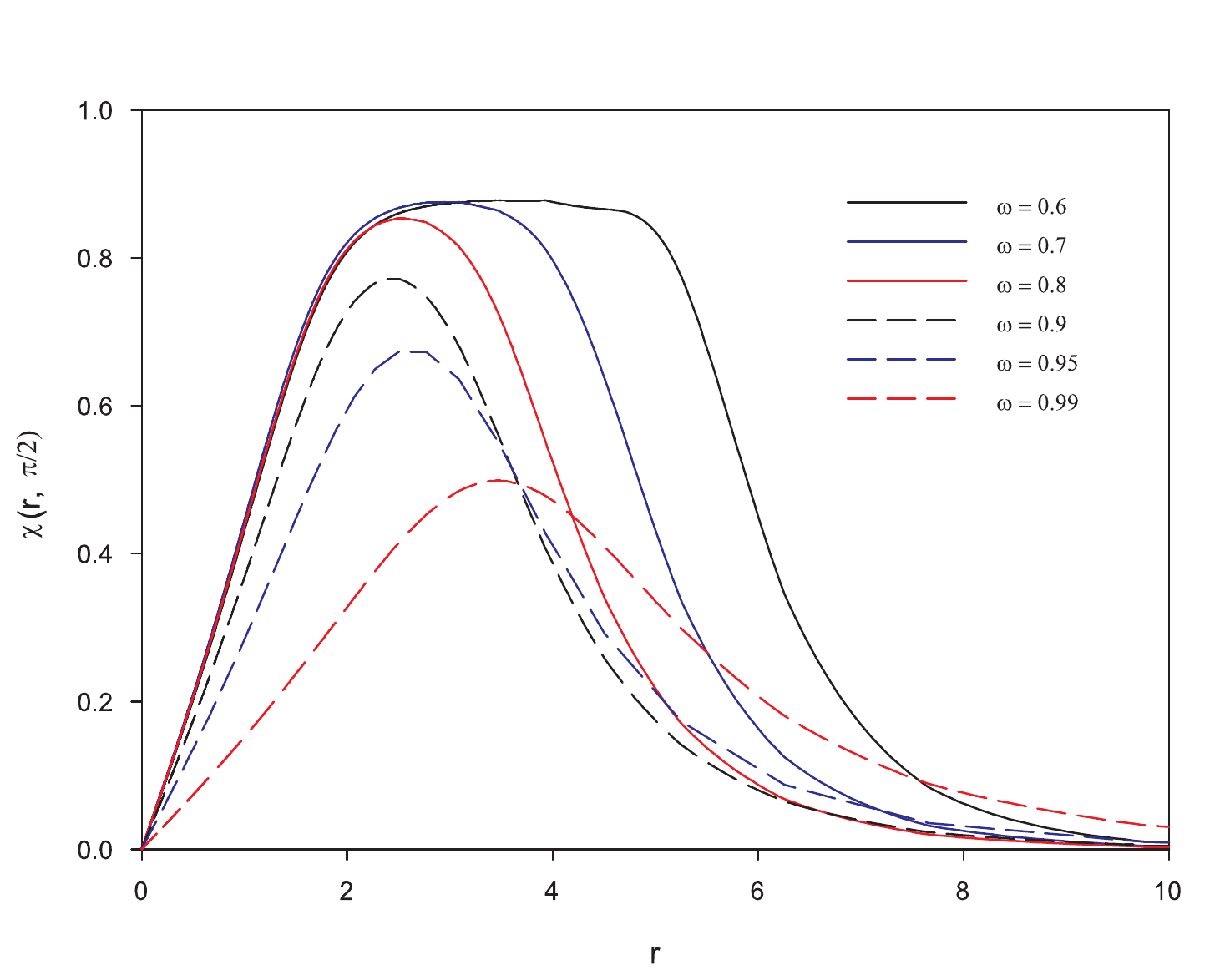}}
   \caption{Left: Energy density in the equator plane for the spinning Q-balls with boundary conditions (\ref{bound1}), $N=10$ and $C=C_{10}=27$ for several values of $\omega$. From top  to bottom: $\omega=0.8$, 0.7, 0.6 (solid lines) and $\omega=0.9$, 0.95, 0.99 (dashed lines). 
            Right: Idem for the field $\chi(r,\pi/2)$ with, from left to right, $\omega=0.8$, 0.7, 0.6 (solid lines) and $\omega=0.9$, 0.95, 0.99 (dashed lines).
         }
\label{t_00_k_11} 
 \end{figure} 
 
A more accurate description of the $\chi$ field and of the energy density in the equator plane is presented in Figs. \ref{t_00_k_11} for several values of $\omega$. We were able to construct solutions for $\omega \in [0.55,0.99]$. The upper limit for $\omega$ is the same as in the spherically symmetric case as it has previously been observed in \cite{kkl}. It is likely that solutions
exist also for smaller $\omega$ but their construction become highly involved and understanding in detail the full pattern of solutions
is not the aim of this exploratory work. The numerical difficulties in obtaining solutions for small $\omega$ are recurrent and also pointed out  in \cite{kkl}. In the case of spherically symmetric solutions, it turns out that several solutions exist with the same $\phi(0)$ and different $\omega$, the values of $\phi(0)$ becoming weakly dependant on $\omega$. In the case of axial solutions, the difficulty seems to be related to the fact that the solution spreads over spacetime and the mesh has to be updated constantly. At least, the spinning solution with the minimal energy occurs inside the interval 
that we explored
and is reached for $\omega \approx 0.86$ where $M\approx 55$ (5.9 GeV)  and $Q \approx 51$. The largest diameter of the deconfined part in this lightest solution is about $8$, that is around $3.5$ fm. 

\subsection{Further axial solution}

The angular dependence of the spherically symmetric and spinning
 solutions discussed above have the same symmetries as the $Y_0^0$ and $Y_1^1 \propto \sin(\theta)$
 spherical harmonics. It is therefore tempting to assume that families of solutions exist with
 the symmetries of the $Y_m^k(\theta, \phi)$
  spherical harmonics with $-m \leq k \leq m$. For instance the parity-odd
 spinning solution of \cite{volkov} has the same angular symmetries as the function $Y_2^1$. According to this observation there
 should exist also within our framework a family of solutions related to \textit{e.g.}
 the function $Y_1^0 \propto \cos(\theta)$.  Such solutions were first constructed in \cite{bh} with $\beta=6$ and different values for the parameter $C$, and here checked to exist for generic values of the potential's parameters.
 
\begin{figure}[ht]
\centering
  \subfigure{\includegraphics[width=7cm]{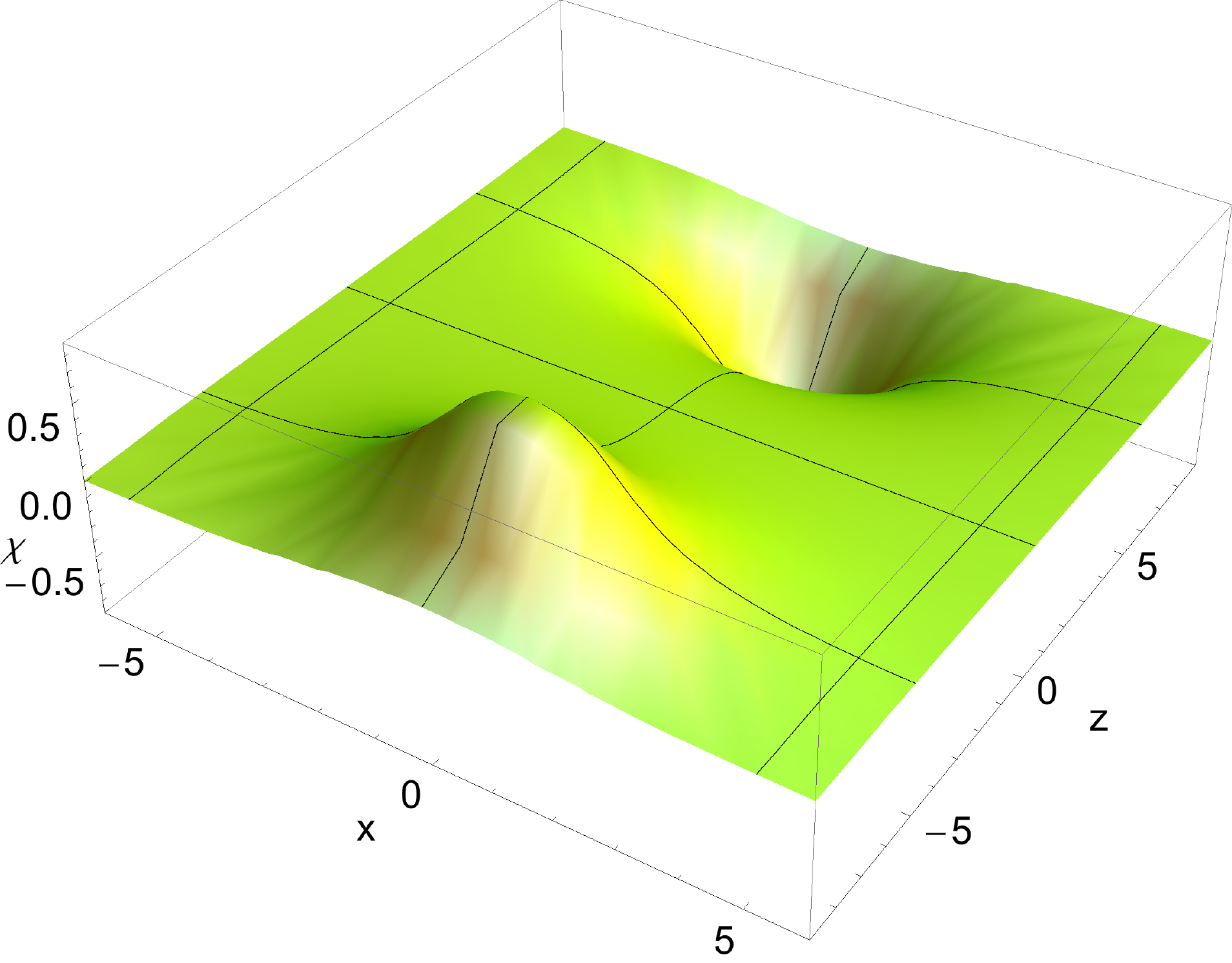}}
    \subfigure{\includegraphics[width=7cm]{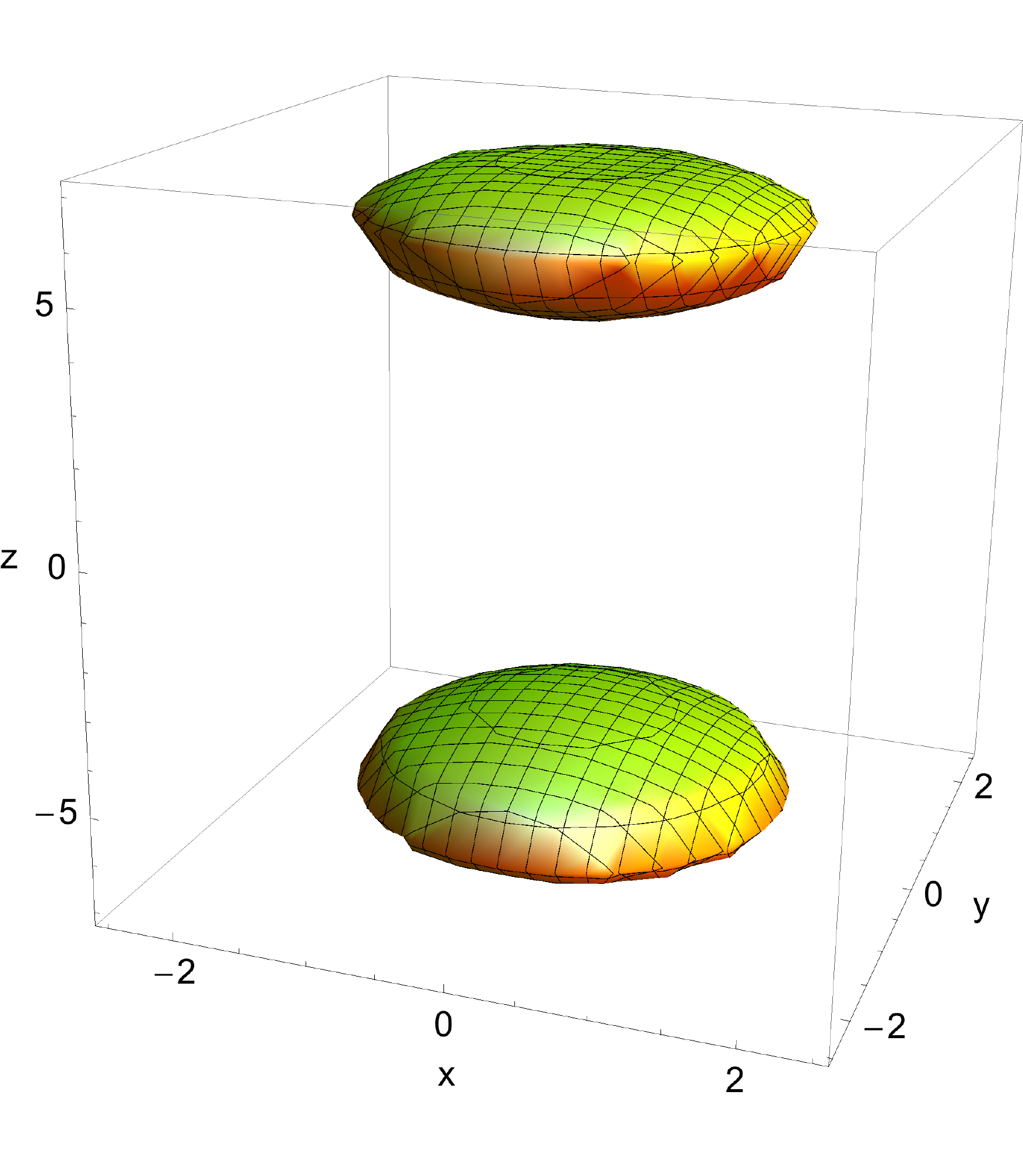}}
\caption{\label{contour_01}
Left: Plot of the scalar field function in the $\phi=0$ plane for the Q-ball with boundary conditions (\ref{bound2}) corresponding to $\beta=10$, $C=C_{10}=27$ and  $\omega=0.8$. Right: Plot of the region where $T_{00}\geq 0.45$ for the same solution. Note that $\{ x,y,z \}$ denote standard cartesian coordinates. }
\end{figure}

Configurations with the $Y_1^0$ angular dependence can be obtained with $\beta=10$ and $C=27$ as Q-ball solutions of Eq. (\ref{pde}) and the boundary conditions
\be \label{bound2}
      \chi(0, \theta) = 0 \ \ , \ \  \chi(\infty, \theta) = 0 \ \ , \ \ \partial \chi(r, \theta=0) = 0 \ \ , \ \
      \chi(r, \theta=\pi/2) = 0.
\ee
These boundary conditions ensure the solution to be odd under parity. We are able to construct with a good accuracy 
the branch of solutions for $\omega \in [0.55,0.95]$. On this interval a configuration with minimal mass
seems to be reached for $\omega \approx 0.92$, corresponding to $M \approx 26$ (2.8 GeV). As an illustration, the solution obtained for $\omega=0.8$ is plotted in Fig. \ref{contour_01}. The norm of the $\chi$-field and the energy density are concentrated in two regions of the $z$-axis, say around $z = \pm z_c$:  We find $z_c \approx 6$ (2.6 fm) for $\omega=0.8$. Such Q-balls could be seen as a two-center deconfined region, much like a bound state of two ``plasma bells". 

\begin{figure}[ht]
\centering
\epsfysize=8cm
\includegraphics[width=10cm]{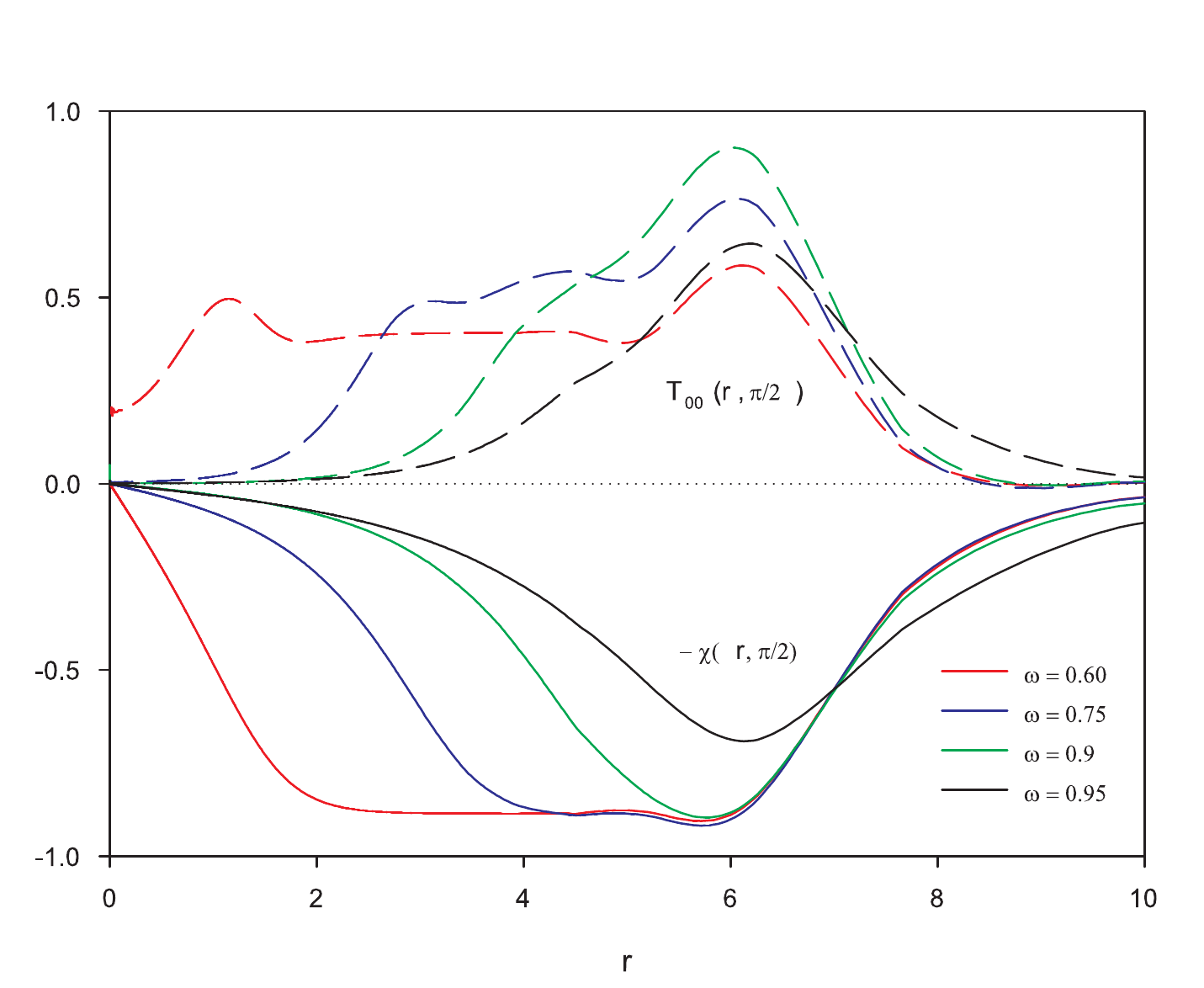}
\caption{\label{T_00_k_10}
Field (solid lines) and energy density (dashed lines) on the positive $z$-axis for the Q-balls with boundary conditions (\ref{bound2}), $\beta=10$ and $C=C_{10}=27$ for different values of $\omega$: From left to right, $\omega=0.6$, 0.75, 0.9, 0.95. The corresponding values on the negative $z$-axis may be obtained by recalling that $\chi$ ($T_{00}$) is an odd (even) function of $z$.}
\end{figure}

The evolution of the energy density on the (positive) $z$-axis for different values of $\omega$ can
be estimated from Fig.  \ref{T_00_k_10}. It suggests that the structure of the lump may become quite involved
when $\omega$ decreases. Again, the construction of such low-$\omega$ Q-balls has not been undertaken, and deconfined regions appear for all the considered values of $\omega$.

\begin{figure}[ht]
  \begin{center}
    \subfigure
    {\label{soliton_a0_e2}\includegraphics[scale=0.55]{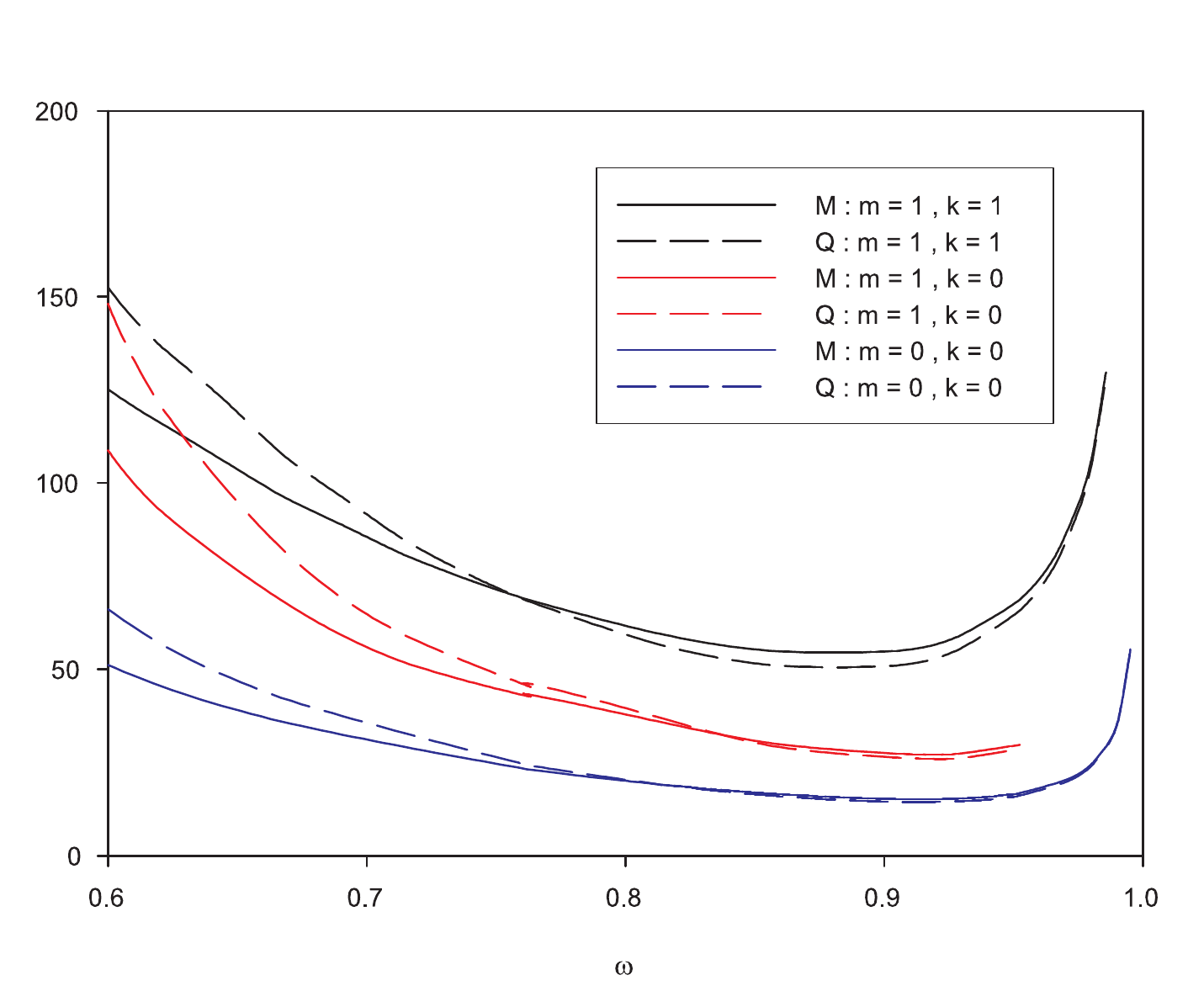}}
    \subfigure
    {\label{soliton_mpsi_e2}\includegraphics[scale=0.55]{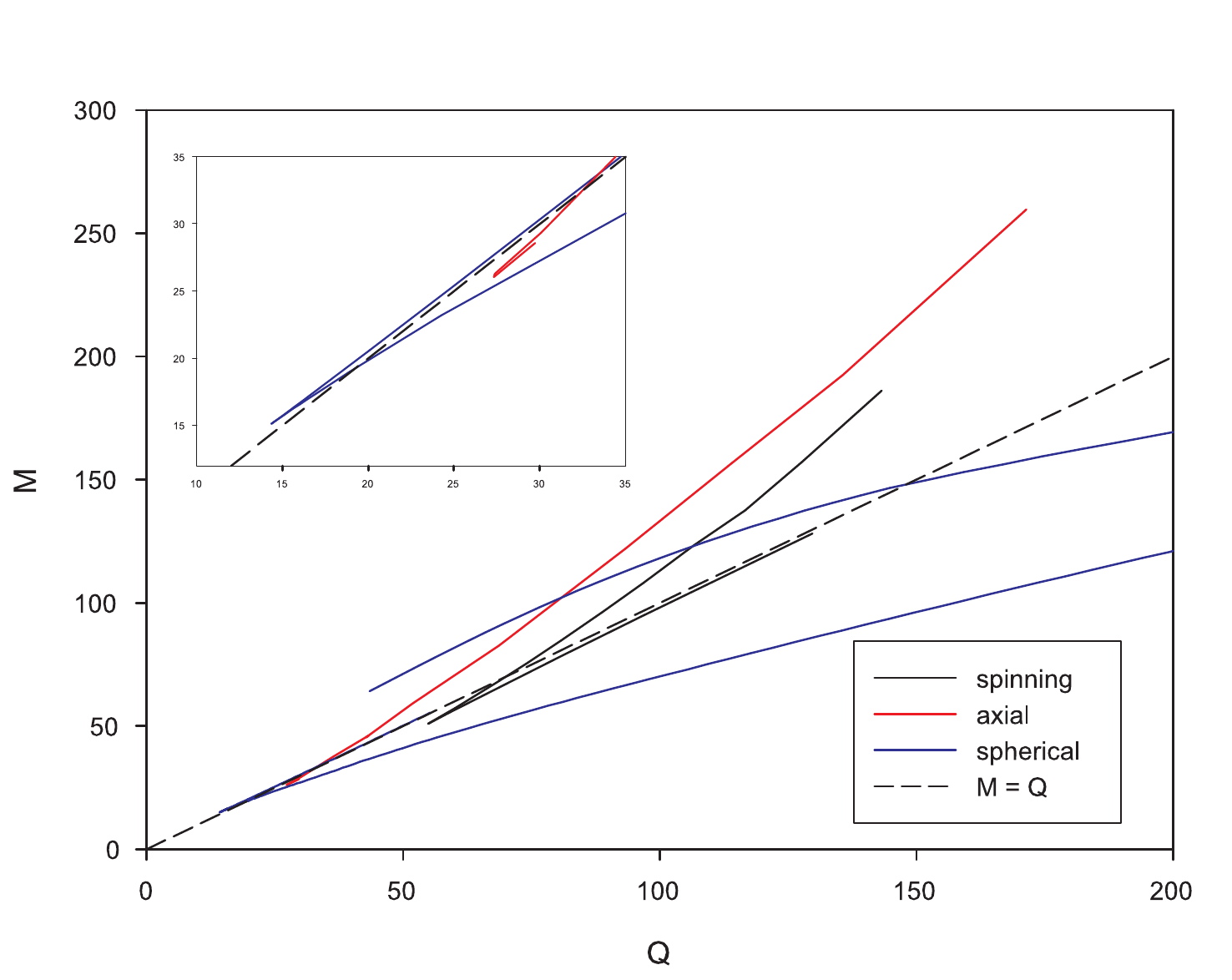}}
\end{center}
   \caption{\label{compara} Left: Mass (solid lines) and charge (dashed lines) as function of $\omega$ of, from top to bottom, the spinning solution (1,1), the axial solution (1,0), and the spherical solution (0,0). Right: Mass versus charge of the different solutions obtained. A zoom on the region with the lowest masses and charges has been also added. 
         }
\label{M_Q_11} 
 \end{figure}  
 
\section{Comparison and stability}\label{compar}

To complete the description of the different solutions just found, we report their mass and charge as functions of the parameter
$\omega$ on the left panel of Fig. 8. This part of the figure is limited to $\omega \in [0.6,0.95]$, where the three types of solutions have been constructed.
Likely due to the large exponent of the field $\beta=10$, the numerical resolution of the partial differential equation (\ref{pde}) becomes unreliable for $\omega < 0.6$;  a more appropriate discretisation of the space of integration
would probably do the job. However, as stated above, we have been able to obtain the solutions with the lowest energy, that are the most likely to be produced at the deconfinement phase transition. 
Let us label the different branches of solutions by the couple $(m,k)$ according to their angular dependence
in relation with the $Y_m^k$ spherical harmonic. As expected, for fixed $\omega$, the solution with the lowest energy
is the spherically symmetric solution, then come successively the axially symmetric $(1,0)$ and the spinning $(1,1)$ solution.

Perhaps more relevant is the right panel of Fig. 8  which represents a $(Q,M)$ plot of the different solutions.
As observed in \cite{kkl} for example, it reveals that, for each family $(m,k)$, the curve $M(Q)$ presents different 
branches terminating into spikes at critical values of the charge.
According to standard arguments based on bifurcation theory,  the branch with the lowest energy should be stable
while the branches with higher energy are unstable, see \textit{e.g.} \cite{kusmartsev}. 

For all values of the conserved charge covered by our solutions, 
the solution with the lowest energy is a spherically
symmetric solution. On the graph, this corresponds to the lowest branch. Note that the two lines corresponding to the spherical solution are plotted up to $Q=200$ on the graph, but actually terminates with a spike at $Q\approx 2200$, $M\approx 806$ (the graph was cut for obvious reason). 

For the values $Q>55$, the two axially symmetric solutions coexist and it can be checked that the spinning solution always possesses the lowest mass at a given charge. 
 In the region $27 < Q < 55$, the spinning solution does not exist and the second lowest mass is 
  the $(1,0)$ axial solution (corresponding to the red line). More details of this branch are shown in the window, which is a zoom of the plot on the low $Q$ and $M$ region. 
Comforting these results about the stability, obtained on the  basis of the 
theory of bifurcations, let us finally point out that the various branches of stable solutions fulfil the 
condition  $M < M_{free}$ , $M_{free} = m_b Q$ where $m_b$ is the mass of the boson in the underlying field
theory (with our units $m_b=1$) and $Q$ is the charge which is, 
in conventional Q-ball literature, is interpreted as the particle number. The Q-balls located under the $M=Q$ line in the right panel of Fig. 8 are then stable. 
 
Several families of solutions could in principle be constructed extending the above results.
Let us mention three possibilities~: (i) radial excitations of the spherically symmetric solution,
where the scalar field presents nodes at some values of the radial variable (see \textit{e.g.} \cite{volkov,kkl});  
(ii) further  angular excitations corresponding to higher
values of the integers $k$ and $m$; (iii) a mix of (i) and (ii).

\section{Finite $N_c$}\label{finite}

The large-$N_c$ limit leads to an effective potential with continuous U(1)-symmetry and consequently to Q-balls. Although a detailed numerical investigation of this case is out the scope of the present paper, it is worth commenting the situation at finite $N_c$. 

When $N_c$ is finite (and $\geq$ 2), the symmetry of the Lagrangian is a discrete, Z$_{N_c}$ one enforced by the potential term $a_{N_c}(\phi^{N_c}+\phi^{*N_c})$, causing the appearance of a term $a_{N_c} N_c \phi^{N_c-1}$ ($a_{N_c} N_c \phi^{*N_c-1}$) in the equation of motion for $\phi$ ($\phi^*$). One readily sees that the ansatz (\ref{ansatz0}) leads to incompatible equations unless $\omega=0$. In this case, the equations of motion (\ref{eom1}) and (\ref{pde}) are recovered. So all the Q-ball solutions with $\omega=0$ existing at infinite $N_c$ are expected to exist at finite $N_c$ too. In  particular, we have found such solutions in the spherically symmetric case, and we have no argument suggesting that axially symmetric solutions with vanishing $\omega$ do not exist, although they are technically complicated to build. 

Which quantum numbers could be used to label these finite-$N_c$ solutions ? The charge $Q$ labelling Q-balls is a consequence of the U(1) symmetry, so it is no more relevant. However, $Q$ is linked to the total angular momentum $J=kQ$ of the solution, which is defined from the energy-momentum tensor, as well as the mass $M$ of the solution. So $M$ and $J$ are still good observables to classify solutions at finite $N_c$, instead of $M$ and $Q$.

The Q-ball-like configurations that one could find at finite $N_c$ are likely to be unstable because of the absence of a conserved charge $Q$. From the paper \cite{Sca01}, we already have evidences that $N_c=3$ configurations of the form $\phi(x^0,\vec x)=a(x^0)$ exist, where $a(x^0)$ start from the ``deconfined minimum" of the potential and falls down to the confined one as time increases. This is an example of a solution modelling an unstable configuration of deconfined matter. So we think that unstable solutions of the type $\phi(x^0,\vec x)$ may exist, such that $\phi(0,\vec x)=\chi(\vec x)$ can be identified with of the aforementioned $\omega=0$ solutions, but with $\phi(\infty,\vec x)=0$.  This kind of solutions could be interpreted as an unstable plasma ball. 

\section{Conclusions and outlook}\label{Conclu}

In this paper, we have used some knowledge of Q-balls developed several years ago in a context
of non-topological solitons within an effective
model attempting to describe qualitatively the deconfining phase transition in the large-$N_c$ limit. The effective theory used is a scalar field theory with U(1)-symmetric potential (seen as the infinite $N_c$ limit of a Z$_{N_c}$--symmetric theory), where the complex scalar field is the color-averaged Polyakov loop. At the deconfinement temperature, the effective potential has two degenerate minima mimicking the first-order deconfinement phase transition. Such an effective potential has a shape that strongly resembles to potentials typically used in the study of Q-balls, motivating the present study.

For technical convenience, the infinite wall of the effective Polyakov loop potential occurring for large norms of the effective field have been replaced here by a large power of the scalar field. As expected, the main families of Q-balls, obtained in previous studies, can be constructed for this potential too. Through an investigation of the energy density distribution in the different Q-balls obtained, we have been able to interpret some of them as bubbles of deconfined matter, or quark-gluon plasma. The different shapes we have found correspond to (by increasing mass for a given charge): spheres, tori, or two-bell bound states. These last two solutions have been obtained here for the first time within a Polyakov-loop model. The physical masses and sizes of the obtained solutions cover a large range of values; we recall that the lightest Q-ball is a spherically symmetric one with a mass $M \approx 1.6$ GeV and and a typical size of 1.5 fm for the deconfined region. Torus- and two-bell like solutions are systematically heavier (and larger) than this solution. 

Returning to potential (\ref{V2}), it could be mentioned that
our Q-balls corresponding to $\omega=0$ could provide solutions of the more realistic, finite-$N_c$, $Z_{N_c}$ equation. We believe that our 
Q-balls with generic values of $\omega$ could produce mean values (or suitable initial profile)
of more realistic time-dependent solutions of the potential (\ref{V2}). As far as the stability of the
solutions is concerned, the non conservation of $Q$ for the realistic equations would allow the lump to decay, as expected in the context of the fireball expansion in hadronic collisions.

The precise behaviour of the solutions in regions of low $\omega$ were not 
pushed to details since it was not the aim of this exploratory study: A more detailed classification of solutions, with higher angular momentum in particular, as well as a more realistic description of the gluon plasma -- through the building of unstable, finite-$N_c$, solutions for example-- is left for future works.


\begin{thebibliography}{99}

\bibitem{Rev} S. Sarkar, H. Satz, and B. Sinha, \textit{The Physics of the Quark-Gluon Plasma} (Springer, 2010). 
\bibitem{Polya} L.~Susskind,
  Phys.\ Rev.\  D {\bf 20}, 2610 (1979); N. Weiss, Phys. Rev. D \textbf{24}, 476 (1981); N. Weiss, Phys. Rev. D \textbf{25}, 2668 (1982).
\bibitem{sve82} B. Svetitsky and L. G. Yaffe, Nucl. Phys. B \textbf{210}, 423
(1982); L. G. Yaffe and B. Svetitsky, Phys. Rev. D \textbf{26}, 963 (1982).
\bibitem{Pisa}
  R.~D.~Pisarski,
  Phys.\ Rev.\  D {\bf 62}, 111501 (2000).
      \bibitem{Dumitru00} 
  A.~Dumitru and R.~D.~Pisarski,
  Phys.\ Lett.\ B {\bf 504}, 282 (2001).
\bibitem{Pisa_app} E.~S.~Fraga, R.~D.~Pisarski and J.~Schaffner-Bielich,
  Phys.\ Rev.\  D {\bf 63}, 121702 (2001); O.~Scavenius, A.~Dumitru and J.~T.~Lenaghan,
  Phys.\ Rev.\  C {\bf 66}, 034903 (2002).
\bibitem{sanni05} F. Sannino, Phys. Rev. D \textbf{72}, 125006 (2005).
\bibitem{Buiss11} 
  F.~Buisseret and G.~Lacroix,
  Phys.\ Rev.\ D {\bf 85}, 016009 (2012).
\bibitem{fukupnjl} K.~Fukushima, Phys.\ Lett.\  B {\bf 591}, 277 (2004); C.~Ratti, M.~A.~Thaler and W.~Weise, Phys.\ Rev.\  D {\bf 73}, 014019 (2006); L.~McLerran, K.~Redlich and C.~Sasaki, Nucl.\ Phys.\  A{\bf 824}, 86 (2009); K.~Fukushima, M.~Ruggieri and R.~Gatto,  Phys.\ Rev.\  D {\bf 81}, 114031 (2010) .
  \bibitem{Sca01} 
  O.~Scavenius, A.~Dumitru and A.~D.~Jackson,
  Phys.\ Rev.\ Lett.\  {\bf 87}, 182302 (2001);   O.~Scavenius, A.~Dumitru, E.~S.~Fraga, J.~T.~Lenaghan and A.~D.~Jackson,
  Phys.\ Rev.\ D {\bf 63}, 116003 (2001);   A.~Dumitru and R.~D.~Pisarski,
  Nucl.\ Phys.\ A {\bf 698}, 444 (2002); E.~S.~Fraga and G.~Krein,
  Phys.\ Lett.\ B {\bf 614}, 181 (2005).
\bibitem{Sca02} 
  O.~Scavenius, A.~Dumitru and J.~T.~Lenaghan,
  Phys.\ Rev.\ C {\bf 66}, 034903 (2002).
  \bibitem{Gupta10} 
  U.~S.~Gupta, R.~K.~Mohapatra, A.~M.~Srivastava and V.~K.~Tiwari,
  Phys.\ Rev.\ D {\bf 82}, 074020 (2010).
\bibitem{Meisinger:2003id} 
  P.~N.~Meisinger, M.~C.~Ogilvie and T.~R.~Miller,
  Phys.\ Lett.\ B {\bf 585}, 149 (2004)
  [hep-ph/0312272].

\bibitem{Sasaki:2012bi} 
  C.~Sasaki and K.~Redlich,
thermodynamics,''
  Phys.\ Rev.\ D {\bf 86}, 014007 (2012)
  [arXiv:1204.4330 [hep-ph]].

 \bibitem{Ruggieri:2012ny} 
   M.~Ruggieri, P.~Alba, P.~Castorina, S.~Plumari, C.~Ratti and V.~Greco,
  Phys.\ Rev.\ D {\bf 86}, 054007 (2012)
  [arXiv:1204.5995 [hep-ph]].

Moreover, descriptions based on AdS/QCD are nowadays available, see for example:

\bibitem{Kajantie:2006hv} 
  K.~Kajantie, T.~Tahkokallio and J.~-T.~Yee,
  JHEP {\bf 0701}, 019 (2007)
  [hep-ph/0609254].

\bibitem{Bigazzi:2011db} 
  F.~Bigazzi, A.~L.~Cotrone, J.~Mas, D.~Mayerson and J.~Tarrio,
  Commun.\ Theor.\ Phys.\  {\bf 57}, 364 (2012)
  [arXiv:1110.1744 [hep-th]].

\bibitem{Kelley:2011ds} 
  T.~M.~Kelley,
  arXiv:1108.0653 [hep-ph].
  
  
  
  \bibitem{coleman} S. Coleman, Nucl. Phys. B {\bf 262}, 263 (1985) (E: B \textbf{269}, 744 (1986)).

\bibitem{lee_pang} T. D. Lee and Y. Pang, Phys. Rept. {\bf 221} 251 (1992).
\bibitem{make} Y. Makeenko, \textit{Methods of contemporary gauge theory}  (Cambridge University Press, 2002).
\bibitem{TcN} B.~Lucini, M.~Teper and U.~Wenger, JHEP {\bf 0502}, 033 (2005).
\bibitem{gupta07} S.~Gupta, K.~Huebner and O.~Kaczmarek, Phys.\ Rev.\  D {\bf 77}, 034503 (2008).
\bibitem{braun}
  J.~Braun, A.~Eichhorn, H.~Gies, J.~M.~Pawlowski,
  Eur.\ Phys.\ J.\  C{\bf 70}, 689 (2010).

  \bibitem{pane09} M.~Panero, Phys.\ Rev.\ Lett.\  {\bf 103}, 232001 (2009).


\bibitem{volkov} M. S. Volkov and E. Wohnert, Phys. Rev. D \textbf{66}, 085003 (2002).
\bibitem{kkl} B. Kleihaus, J. Kunz and M. List, Phys. Rev. D \textbf{72}, 064002 (2005).  
\bibitem{colsys}
U. Ascher, J. Christiansen and R. D. Russell, Math. Comput. {\bf 33}
(1979), 659; ACM Trans. Math. Softw. {\bf 7} (1981), 209.

\bibitem{Arodz:2009ye} 
  H.~Arodz, J.~Karkowski and Z.~Swierczynski,
  Phys.\ Rev.\ D {\bf 80}, 067702 (2009)
  [arXiv:0907.2801 [hep-th]].

\bibitem{Brihaye:2009dx} 
  Y.~Brihaye, T.~Caebergs and T.~Delsate,
  arXiv:0907.0913 [gr-qc].

\bibitem{Campanelli:2009su} 
  L.~Campanelli and M.~Ruggieri,
  Phys.\ Rev.\ D {\bf 80}, 036006 (2009)
  [arXiv:0904.4802 [hep-th]].

\bibitem{fidisol} W. Schonauer and R. Weiss, J. Comput. Appl. Math. {\bf 27}, 279 (1989).\\
M. Schauder, R. Weiss and W. Schonauer, \textit{The CADSOL Program Package}, Universitat Karlsruhe,
Interner Bericht Nr. 46/92 (1992).


\bibitem{bh}  Y. Brihaye and B. Hartmann, Nonlinearity \textbf{21}, 1937 (2008).

\bibitem{kacz} O.~Kaczmarek and F.~Zantow, Phys.\ Rev.\  D {\bf 71}, 114510 (2005).
\bibitem{kusmartsev} F. V. Kusmartsev, Phys. Rep. C {\bf 183}, 1 (1989).
\end{thebibliography}
\end{document}